\documentclass[11pt]{cernrep}
\usepackage{epsfig}
\usepackage{graphicx}
\usepackage{cite,./mcite}

\begin{document}
\title{Including heavy quark production in PDF fits \\
HERA - LHC Workshop 2007}
\author{Amanda Cooper-Sarkar}
\institute{Oxford University}
\maketitle
\begin{abstract}
At HERA heavy quarks may contribute up to $30\%$ of the structure function 
$F_2$. The introduction of heavy quarks requires an extension of the DGLAP 
formalism. The effect of using different heavy flavour number schemes, and 
different approaches to the running of $\alpha_s$, are compared using
the ZEUS PDF fit formalism. The potential of including charm data in the fit 
is explored, using $D^*$ double differential cross-sections 
rather than the inclusive quantity $F_2^{c\bar{c}}$.  
\end{abstract}


Parton Density Function (PDF) determinations are usually global 
fits~\cite{mrst,mrst06,cteq,zeus-s}, which use inclusive cross-section data and 
structure function measurements from deep inelastic lepton hadron scattering 
(DIS) data as well as some other exclusive cross-ections. 
The kinematics
of lepton hadron scattering is described in terms of the variables $Q^2$, the
invariant mass of the exchanged vector boson, Bjorken $x$, the fraction
of the momentum of the incoming nucleon taken by the struck quark (in the 
quark-parton model), and $y$ which measures the energy transfer between the
lepton and hadron systems.
The differential cross-section for the neutral current (NC) 
process is given in terms of the structure functions by
\[
\frac {d^2\sigma(e^{\pm}p) } {dxdQ^2} =  \frac {2\pi\alpha^2} {Q^4 x}
\left[Y_+\,F_2(x,Q^2) - y^2 \,F_L(x,Q^2)
\mp Y_-\, xF_3(x,Q^2) \right],
\]
where $\displaystyle Y_\pm=1\pm(1-y)^2$. In the HERA kinematic range there 
is a sizeable 
contribution to the $F_2$ structure function from heavy quarks, 
particularly charm. Thus heavy quarks must be properly treated in the fomalism.
 Furthermore fitting data on charm production may help to give constraints 
on the gluon PDF at low-$x$.

The most frequent approaches to the inclusion of heavy quarks
 within the conventional framework of QCD evolution 
using the DGLAP equations~\cite{ap,gl,l,d}  are~\footnote{Charm production is described 
here but a similar formalism describes beauty production}:
\begin{itemize}
\item ZM-VFN (zero-mass variable flavour number schemes) in which the
charm parton density $c(x,Q^2)$ satisfies
$c(x,Q^2)=0$ for $Q^2\leq \mu^2_c$ and $n_f=3+\theta(Q^2-\mu^2_c)$
in the splitting functions and $\beta$ function. The threshold $\mu^2_c$,
which is in the range $m^2_c<\mu^2_c<4m^2_c$, is chosen
so that $F_2^c(x,Q^2)=2e_c^2xc(x,Q^2)$ gives a satisfactory description
of the data. The advantage of this approach is that the simplicity
of the massless DGLAP equations is retained. The disadvantage is that the
physical threshold $\hat{W}^2=Q^2({1\over z}-1)\geq 4m_c^2$ is not treated
correctly ($\hat{W}$ is the $\gamma^*g$ CM energy).
\item FFN (fixed flavour number schemes) in which there is no charm parton
density and all charmed quarks are generated by the BGF process. The 
advantage of the FFNS scheme is that the threshold region is correctly
handled, but the disadvantge is that large $\ln(Q^2/m^2_c)$ terms appear
and charm has to be treated ab initio in each hard process.
\item GM-VFN (general mass variable flavour number schemes), which aim to 
treat the 
threshold correctly and absorb $\ln(Q^2/m^2_c)$ terms into a charm parton
density at large $Q^2$. There are differing versions of such 
schemes~\cite{cteq65,hqnew}
\end{itemize}

\begin{figure}[tbp]
\vspace{-1.0cm} 
\centerline{
\epsfig{figure=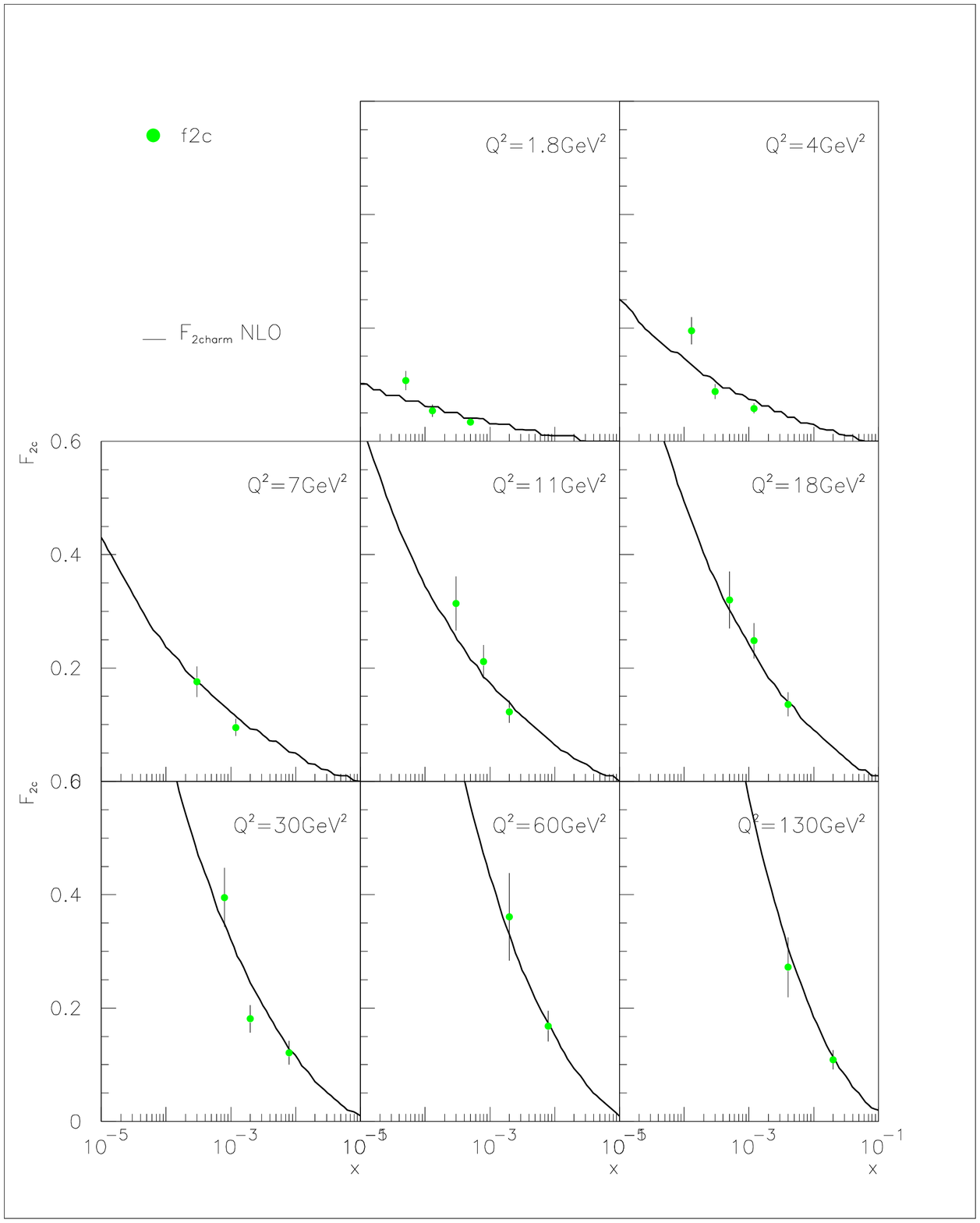,width=0.33\textwidth,height=5cm}
\epsfig{figure=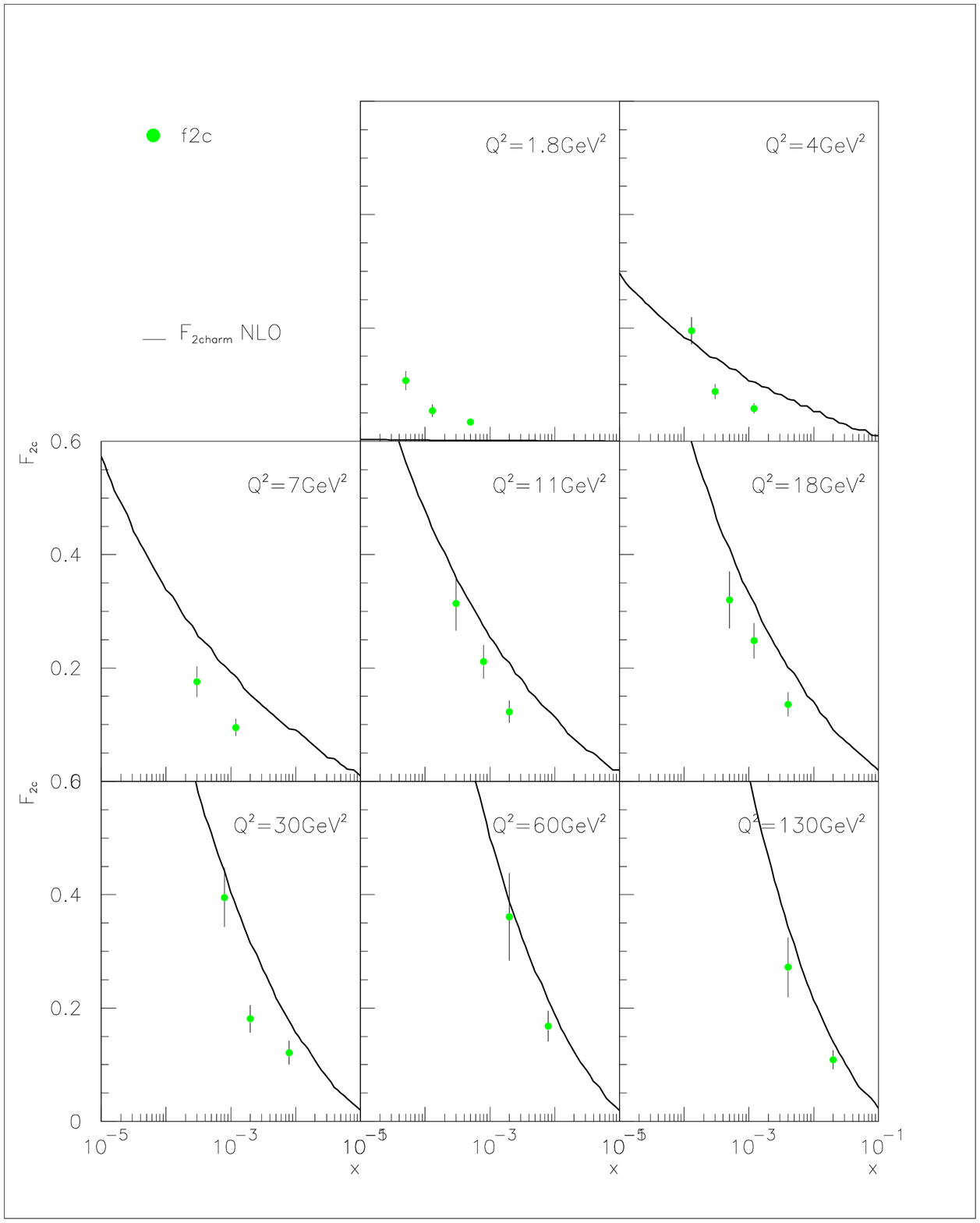,width=0.33\textwidth,height=5cm}
\epsfig{figure=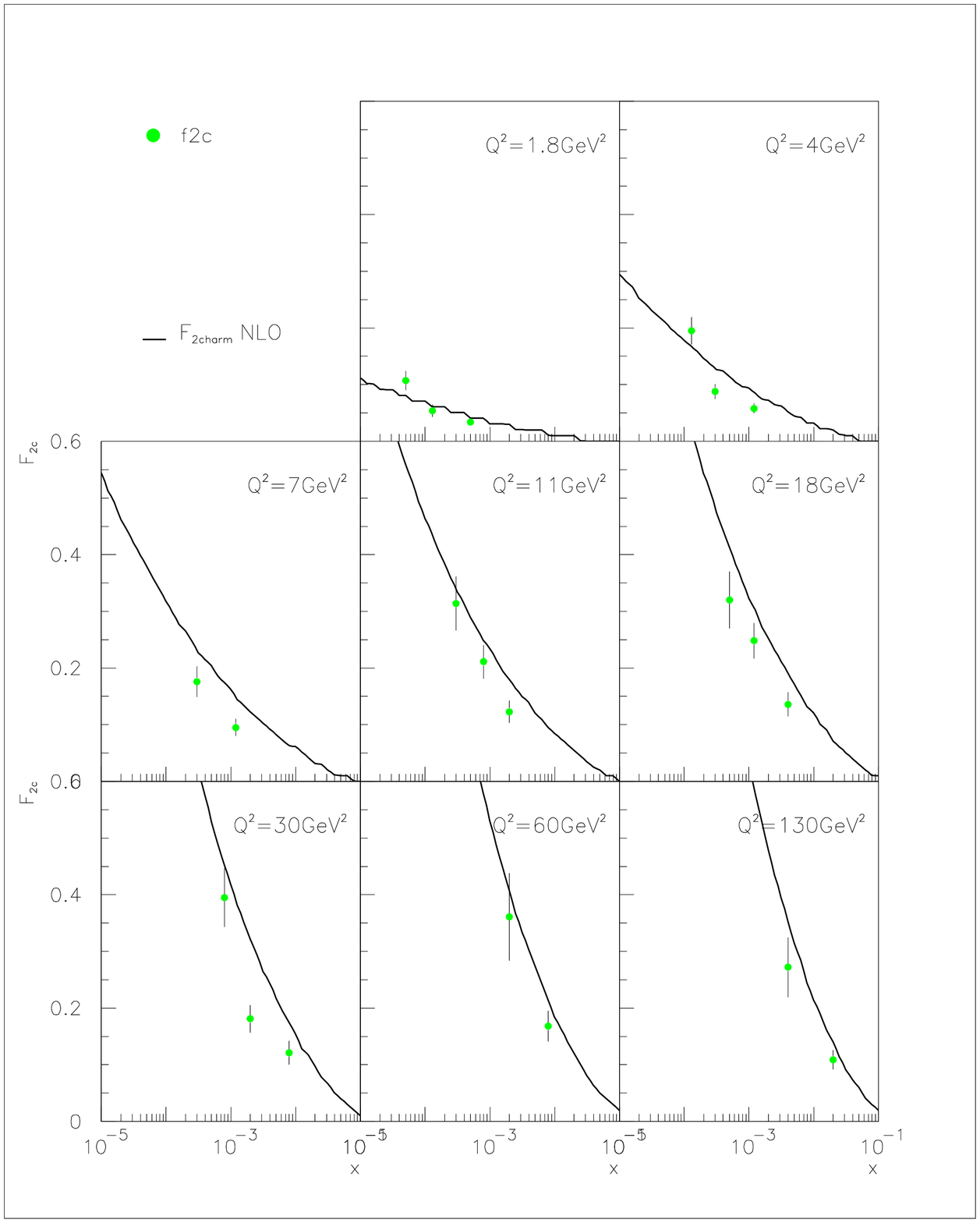,width=0.33\textwidth,height=5cm}}
\vspace{-0.5cm}
\caption {ZEUS data on $F_2^{c\bar{c}}$ compared to predictions using the FFN 
(left) ZM-VFN(middle), TR-VFN (right) schemes. In each case the fit parameters 
are kept the same (fitted using FFN) and only the scheme is changed.}
\label{fig:schemes}
\end{figure}

For the main ZEUS-S analysis~\cite{zeus-s}, 
the heavy quark production scheme used was the
general mass variable flavour number scheme of Roberts and Thorne 
(TR-VFN)~\cite{hq,hqnew}. However we also investigated the use of the FFN 
for 3-flavours and the ZM-VFN.
In Fig.~\ref{fig:schemes} we compare the fit prediction for $F_2^{c\bar{c}}$ 
using each of these schemes to data on $F_2^{c\bar{c}}$ from the ZEUS collaboration~\cite{zcharmearly}.
One can see the differences between the FFN and the ZM-VFN at threshold where 
the ZM-VFN is clearly inadequate. In this kinematic region the TR-VFN is more 
like the FFN. However, the TR-VFN scheme becomes
 more like the ZM-VFN scheme for $Q^2 >> m_c^2$. 

\begin{figure}[tbp]
\vspace{-0.5cm} 
\centerline{
\epsfig{figure=f2c_h.eps,width=0.33\textwidth,height=5cm}
\epsfig{figure=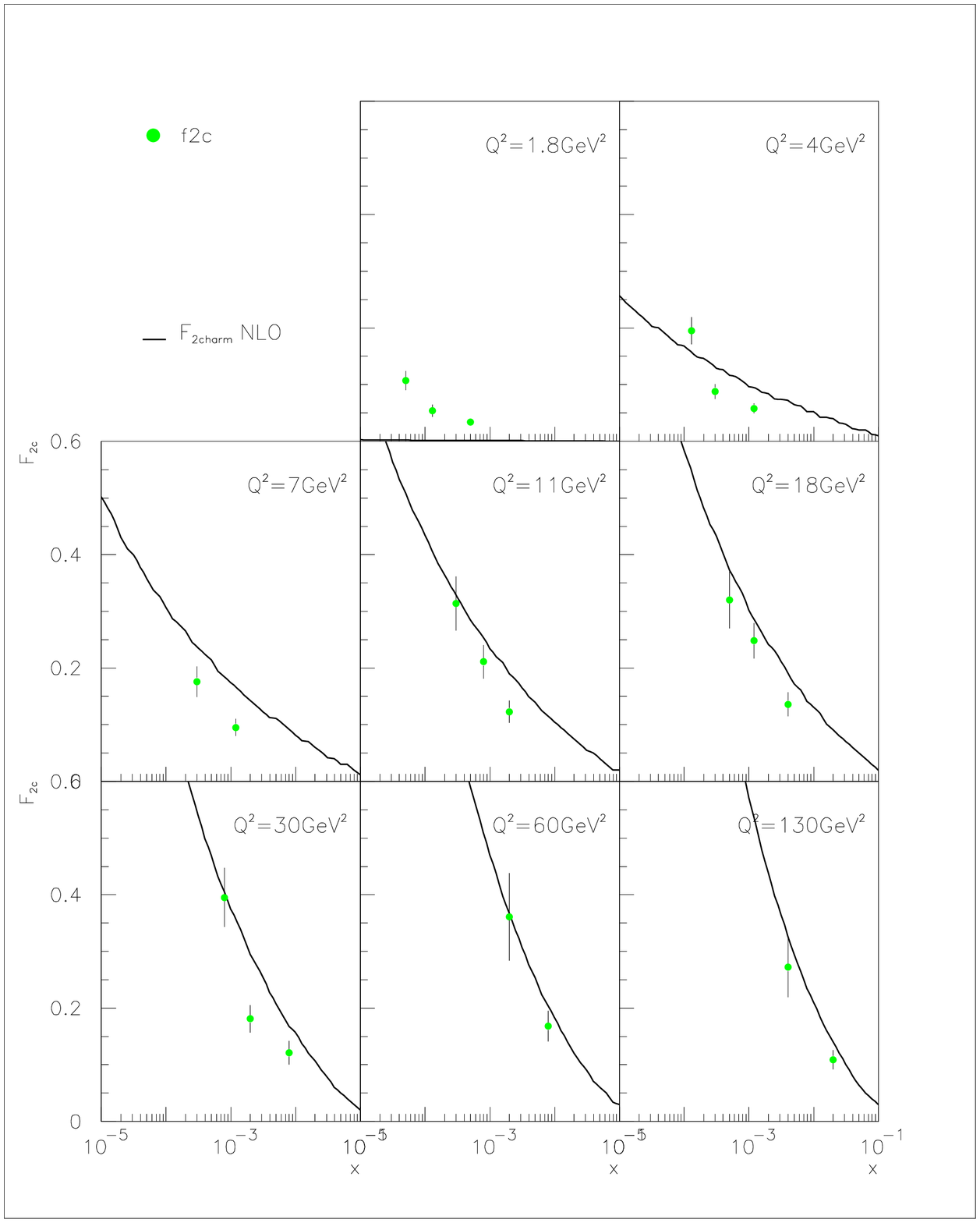,width=0.33\textwidth,height=5cm}
\epsfig{figure=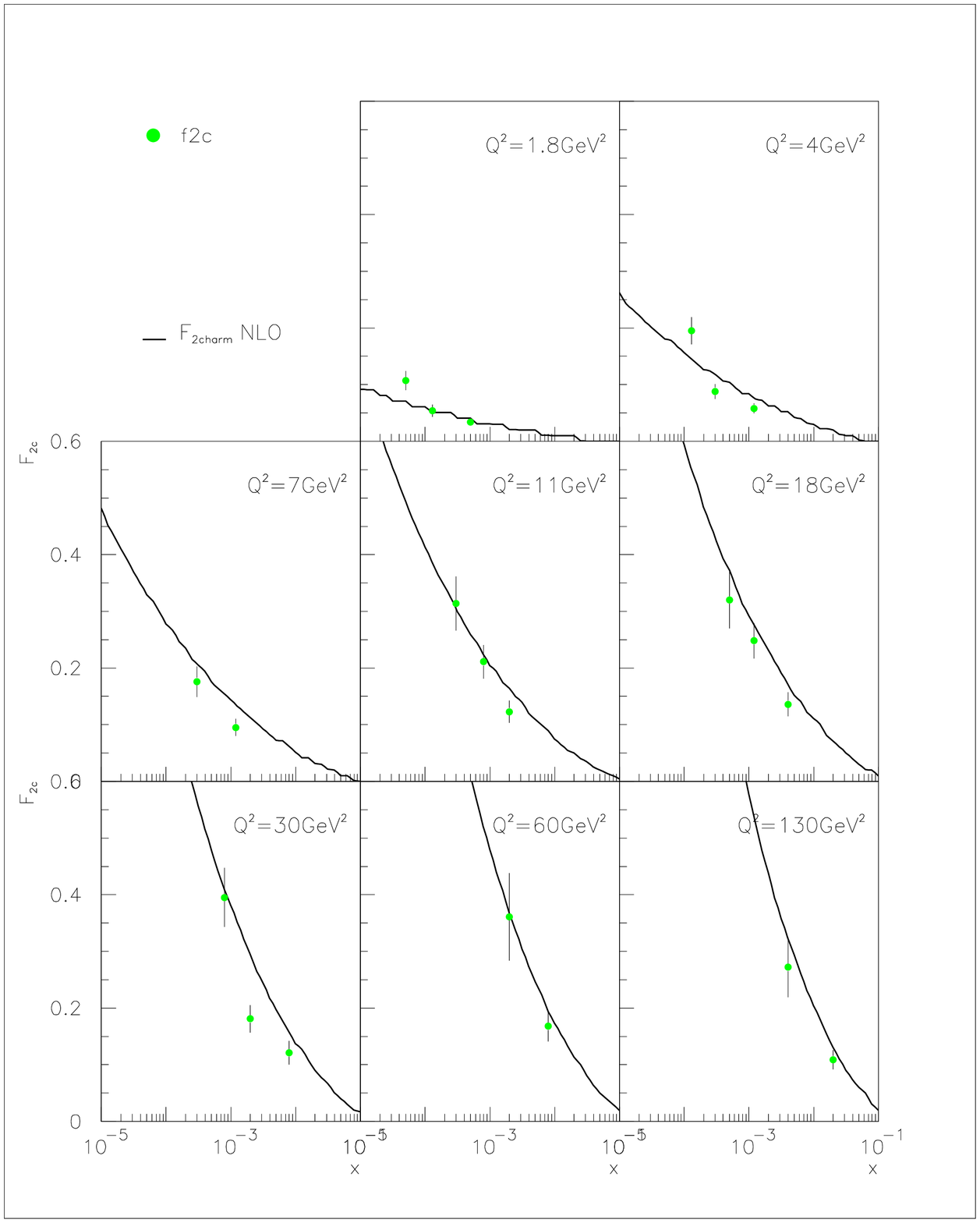,width=0.33\textwidth,height=5cm}}
\vspace{-0.5cm}
\caption {ZEUS data on $F_2^{c\bar{c}}$ compared to predictions using the FFN 
(left) ZM-VFN(middle), TR-VFN (right) schemes. In each case the fit parameters 
are refitted when the scheme is changed.}
\label{fig:fitschemes}
\end{figure}

\begin{figure}[tbp]
\vspace{-0.75cm} 
\centerline{
\epsfig{figure=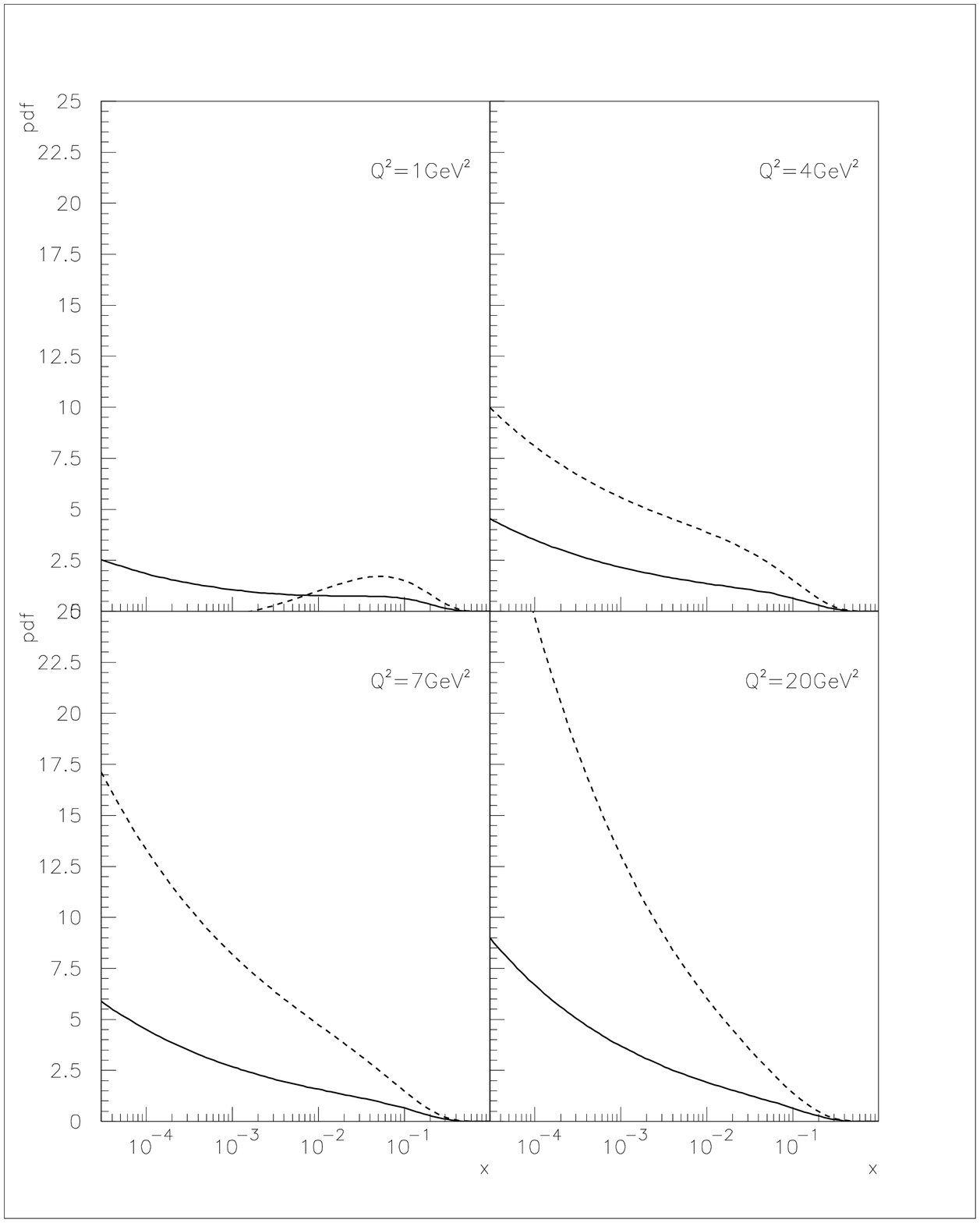,width=0.33\textwidth,height=5cm}
\epsfig{figure=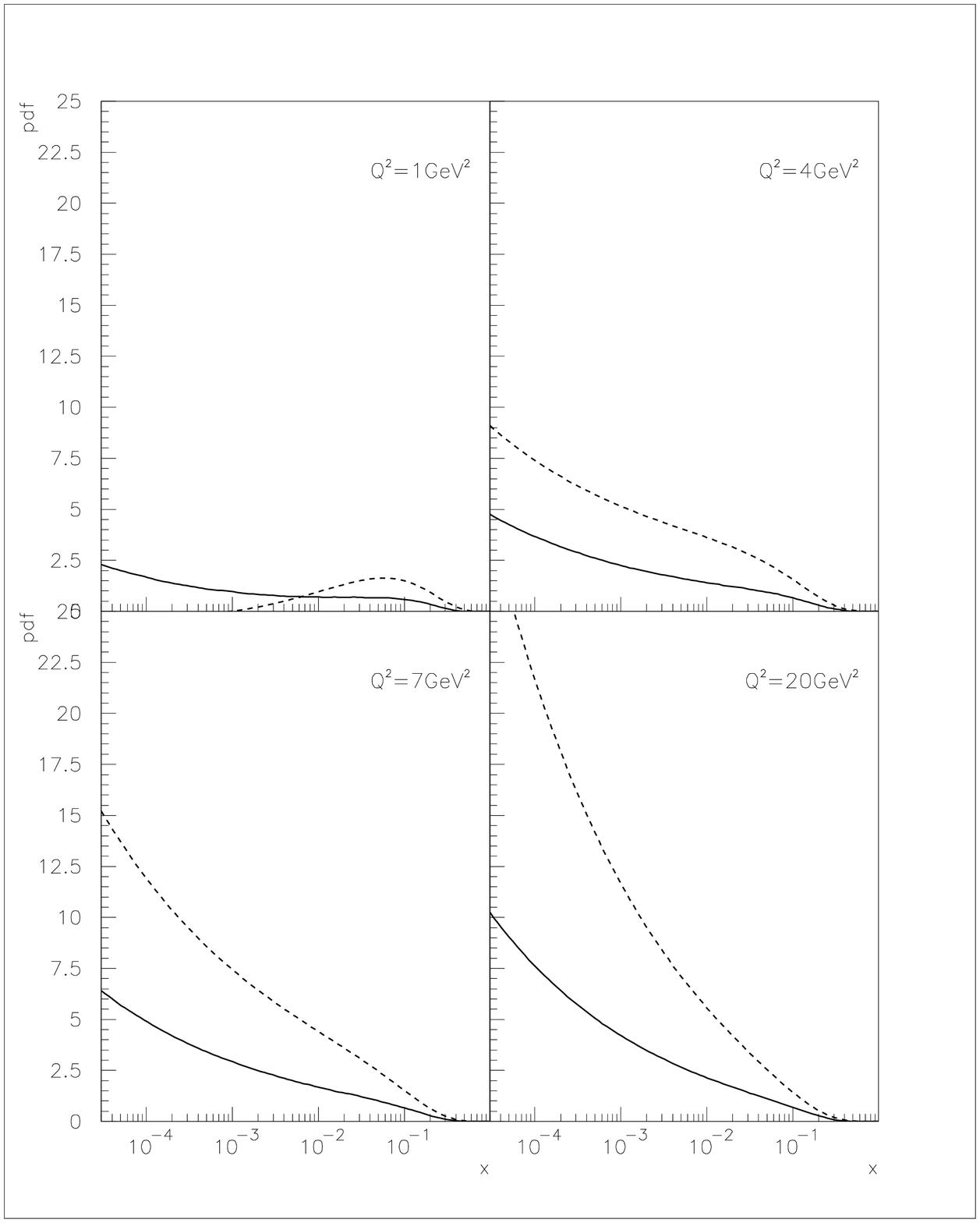,width=0.33\textwidth,height=5cm}
\epsfig{figure=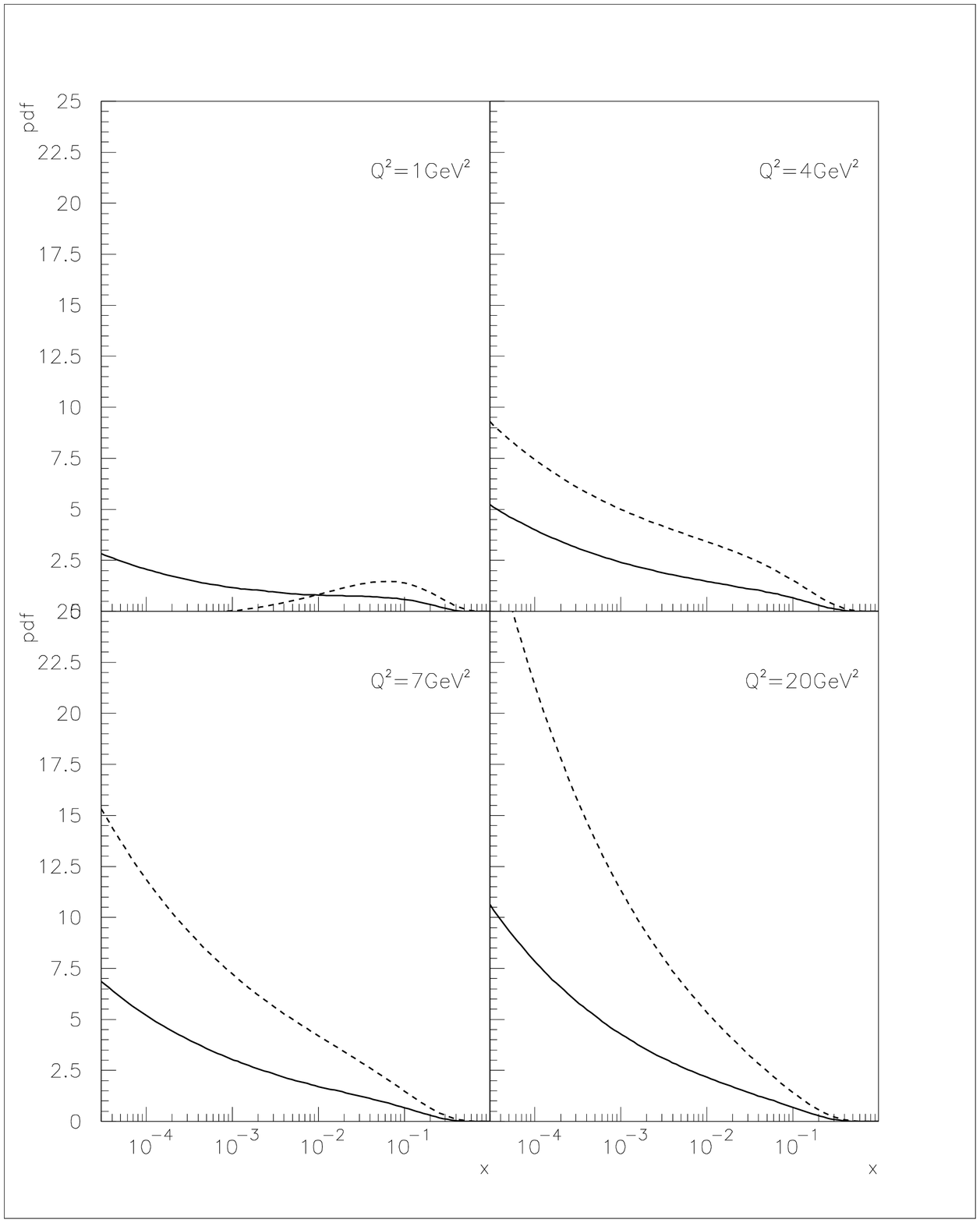,width=0.33\textwidth,height=5cm}}
\caption {The sea and gluon PDFs extracted from fits using the FFN 
(left) ZM-VFN(middle), TR-VFN (right) schemes. In each case the fit parameters 
are refitted when the scheme is changed}
\label{fig:glusea_schemes}
\end{figure}

This comparison illustrates the effect of change in scheme when 
keeping the PDF parameters fixed. In practice one should refit the PDF 
parameters using the alternative schemes. The result of this is shown in 
Fig~\ref{fig:fitschemes}. The difference between the FFN and TR-VFN is not so 
marked.
It is well known that these choices have some effect on the steepness of 
the gluon at very small-$x$, such that the zero-mass choice produces a 
slightly less steep gluon. In Fig~\ref{fig:glusea_schemes} the differing 
shapes of the sea and the gluon PDFs for these different heavy quark 
schemes are illustrated.

\begin{figure}[tbp]
\vspace{-0.5cm} 
\centerline{
\epsfig{figure=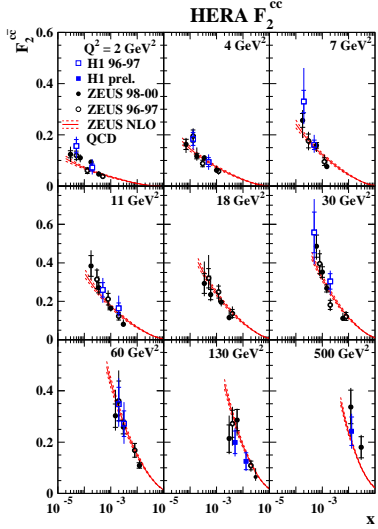,height=7cm}
}
\caption {Comparsion of ZEUS PDF fit predictions to recent charm data from ZEUS and H1 on $F_2^{c\bar{c}}$ }
\label{fig:recentcharm}
\end{figure}

Figure~\ref{fig:recentcharm} shows the ZEUS-S fit 
predictions for more recent $F_2^{c\bar{c}}$ data from ZEUS and 
H1~\cite{dstar2003,h1newcharm}. The scheme 
chosen was FFN for 3 flavours with the renormalisation and factorisation scale 
for light quarks both set to $Q^2$ but the factorisation scale for heavy 
quarks set to $Q^2 + 4 m_c^2$. The reason for these choices of scheme and scale
 is that these are the choices made in the programme HVQDIS~\cite{hvqdis1,hvqdis2,hvqdis3} which was used 
to extract $F_2^{c\bar{c}}$ from data on $D^*$ production. The scale 
choice does not make any signficant difference to the predictions (see later).

\begin{figure}[tbp]
\vspace{1.0cm} 
\centerline{
\epsfig{figure=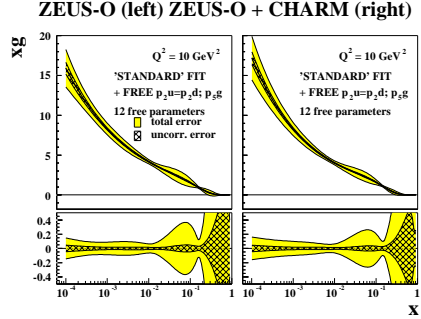,height=8cm}
}
\vspace{-5.0cm}
\caption {The gluon PDF and its fractional uncertainties at $Q^2= 10$GeV$^2$, 
from a) the ZEUS-O PDF fit (left) and b) a smilar fit with $F_2^{c\bar{c}}$ 
data included (right).}
\label{fig:f2charm}
\end{figure}

Note that in Fig~\ref{fig:recentcharm}
 the charm data are shown compared to the ZEUS-S PDF fit 
predictions but these data were not input to the fit. Including the 
ZEUS charm data~\cite{dstar2003} in the ZEUS-S PDF fit gives no 
visible improvement to PDF uncertainties. To investigate the potential of 
charm data to constrain the gluon PDF, we modified the ZEUS-S PDF
fit as follows: all ZEUS inclusive neutral current and charged current cross-section data from HERA-I was included 
but no fixed target data; the parametrisation was modified to free the mid-$x$ 
gluon parameter $p_5(g)$ and the low-$x$ valence parameter $p_2(u)= p_2(d)$, however 
the $\bar{d}-\bar{u}$ normalisation had to be fixed since 
there is no information on this without fixed taregt data. 
This fit is called the ZEUS-O fit. 
Fig.~\ref{fig:f2charm} compares the gluon PDF and its uncertainties as 
extracted from this ZEUS-O PDF fit  
with the those extracted from a similar fit  
including the  $F_2^{c\bar{c}}$ data.
This illustrates that the charm data has the potential to constrain the 
gluon PDF uncertainties. Its lack of impact on the global fit may be because 
we are not using the charm data optimally.
 
\begin{figure}[tbp]
\vspace{-0.5cm} 
\centerline{
\epsfig{figure=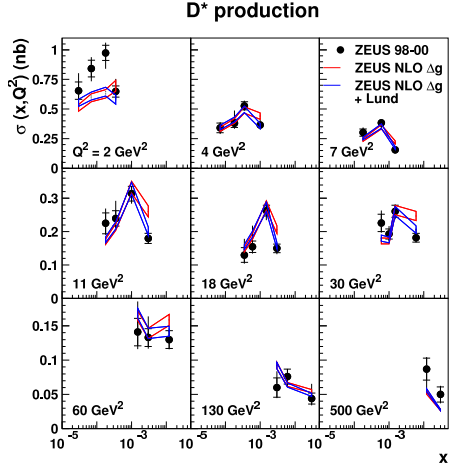,height=8cm}
}
\vspace{-2.0cm}
\caption {Double differential cross-sections for $D^*$ production. The red lines
 show the predictions of the ZEUS-S-13 NLO PDF fit using the Petersen 
fragmentation function for the $D^*$, whereas the blue lines show these 
predictions using the Lund fragmentation function.}
\label{fig:9xsecns}
\end{figure}

$F_2^{c\bar{c}}$ is a quantity extracted from $D^*$ cross-sections by quite a 
large extrapolation. It would be better to fit to those cross-sections 
directly. The evaluation of the theoretical predictions involves running the 
NLO programme HVQDIS for each iteration of the fit. However, one can shorten 
this process by using the 
same method as was used for the ZEUS-JETS fit~\cite{zeusj}.  The
PDF independent subprocess cross-sections are output onto a grid, 
such that they can simply 
be multiplied by the PDFs at each iteration. 
The data used are the nine double differential cross-section measurements of
$d^2\sigma(D^*)/dQ^2dy$~\cite{dstar2003}, see Fig.~\ref{fig:9xsecns}

There are further theoretical 
considerations to be accounted for when unputting $D^*$ cross-sections, 
as opposed to and inclusive quantity like $F_2^{c\bar{c}}$, to a PDF fit. 
Since the grids are calculated using HVQDIS the fit must use the FFN scheme to 
be compatible. This means that we cannot use ZEUS high-$Q^2$ data, since this 
scheme is not suitable at high-$Q^2$. Hence we chose to use the ZEUS-S global 
fit, which incuded fixed target data, with a cut-off $Q^2 < 3000$GeV$^2$. 
Furthermore, it has only recently become evident that since we are using 
the FFN scheme we must also treat the running of $\alpha_s$ differently than in
the VFN schemes. In these VFN schemes $\alpha_s$ is matched at flavour 
threholds~\cite{match}, but the slope of $\alpha_s$ is discontinuous at 
the flavour thresholds. For consistency with HVQDIS we must use 
a 3-flavour $\alpha_S$ which is continuous in $Q^2$. This requires 
an equivalent value of $\alpha_s(M_Z)=0.105$ in order to be consistent, 
at low $Q^2$, with the results of using a value of $\alpha_s(M_Z)=0.118$ in 
the usual VFN schemes. Such a 3-flavour $\alpha_S$ has also been used in 
specialised PDF fits of MRST (MRST2004F3)~\cite{mrstff04}, which are used to make predictions 
for charm production.

\begin{figure}[tbp]
\vspace{-0.5cm} 
\centerline{
\epsfig{figure=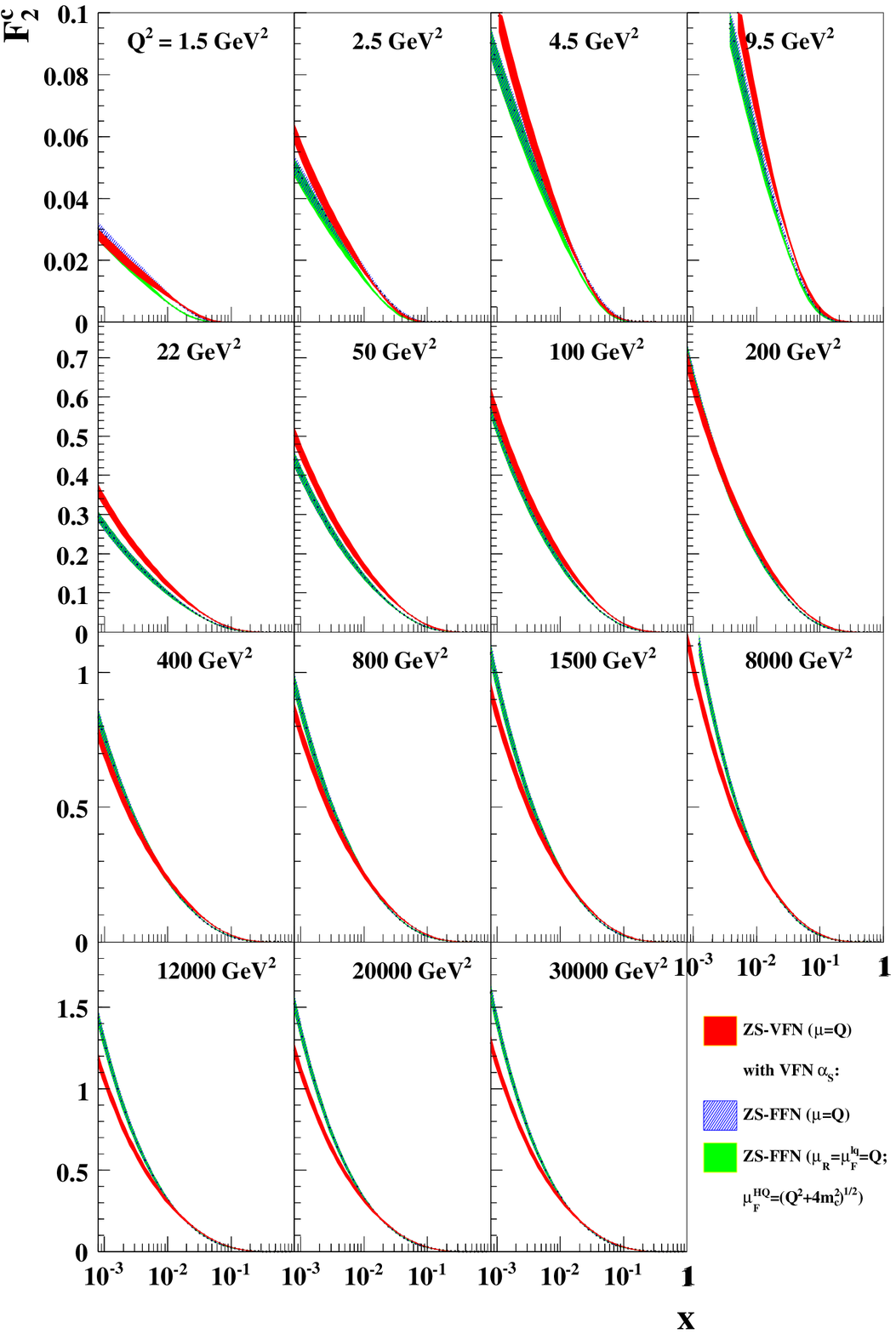,width=0.5\textwidth,height=6cm}
\epsfig{figure=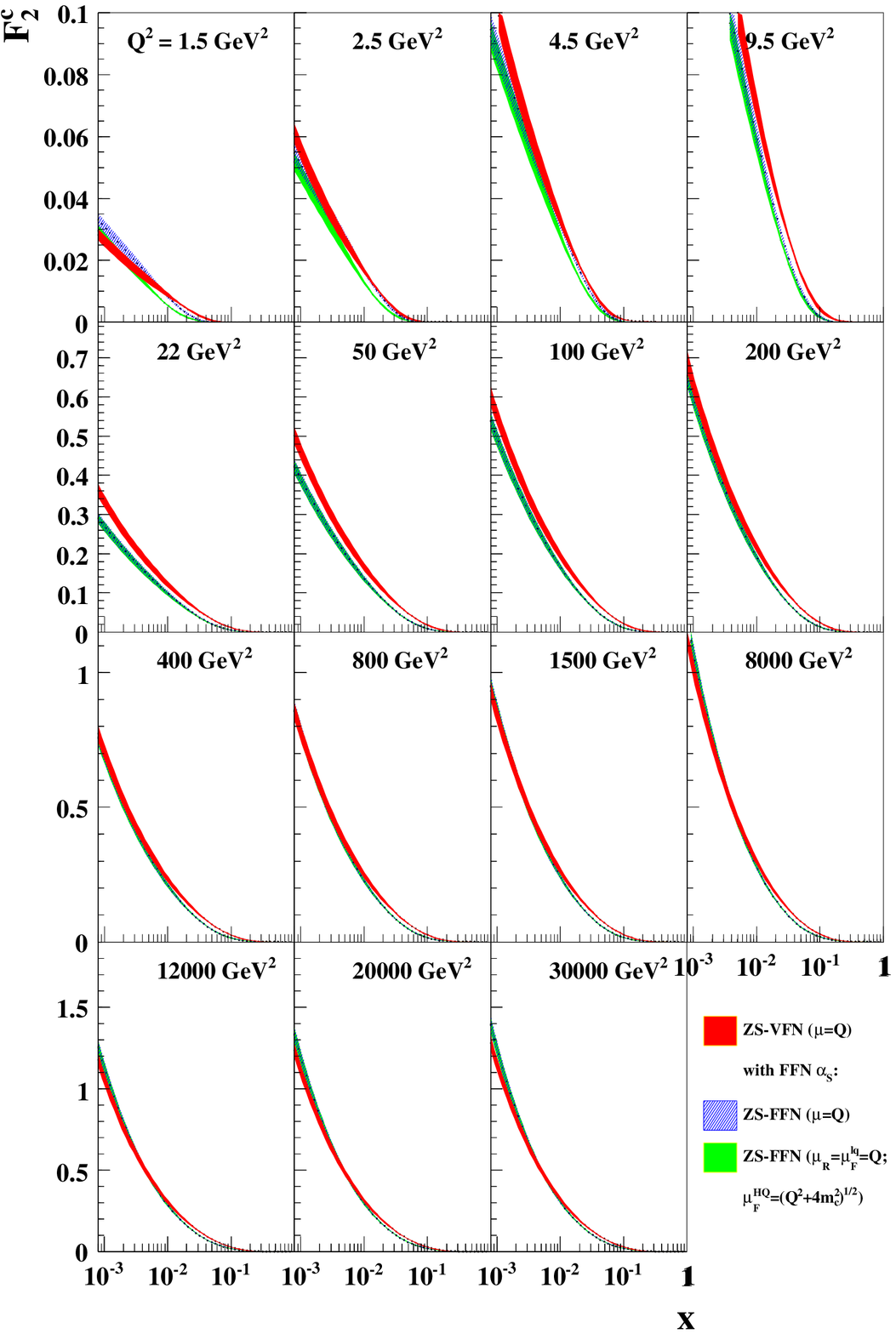,width=0.5\textwidth,height=6cm}}
\caption { Comparison of predictions for $F_2^{c\bar{c}}$, from fits which 
use the TR-VFN scheme and the FFN scheme with two different factorisation 
scales: on the left hand side the FFN schemes still use a VFN treatment of
$\alpha_s$, whereas on the right hand side a 3-flavour $\alpha_s$ is used.}
\label{fig:f2calphas}
\end{figure}

\begin{figure}[tbp]
\vspace{-1.5cm} 
\centerline{
\epsfig{figure=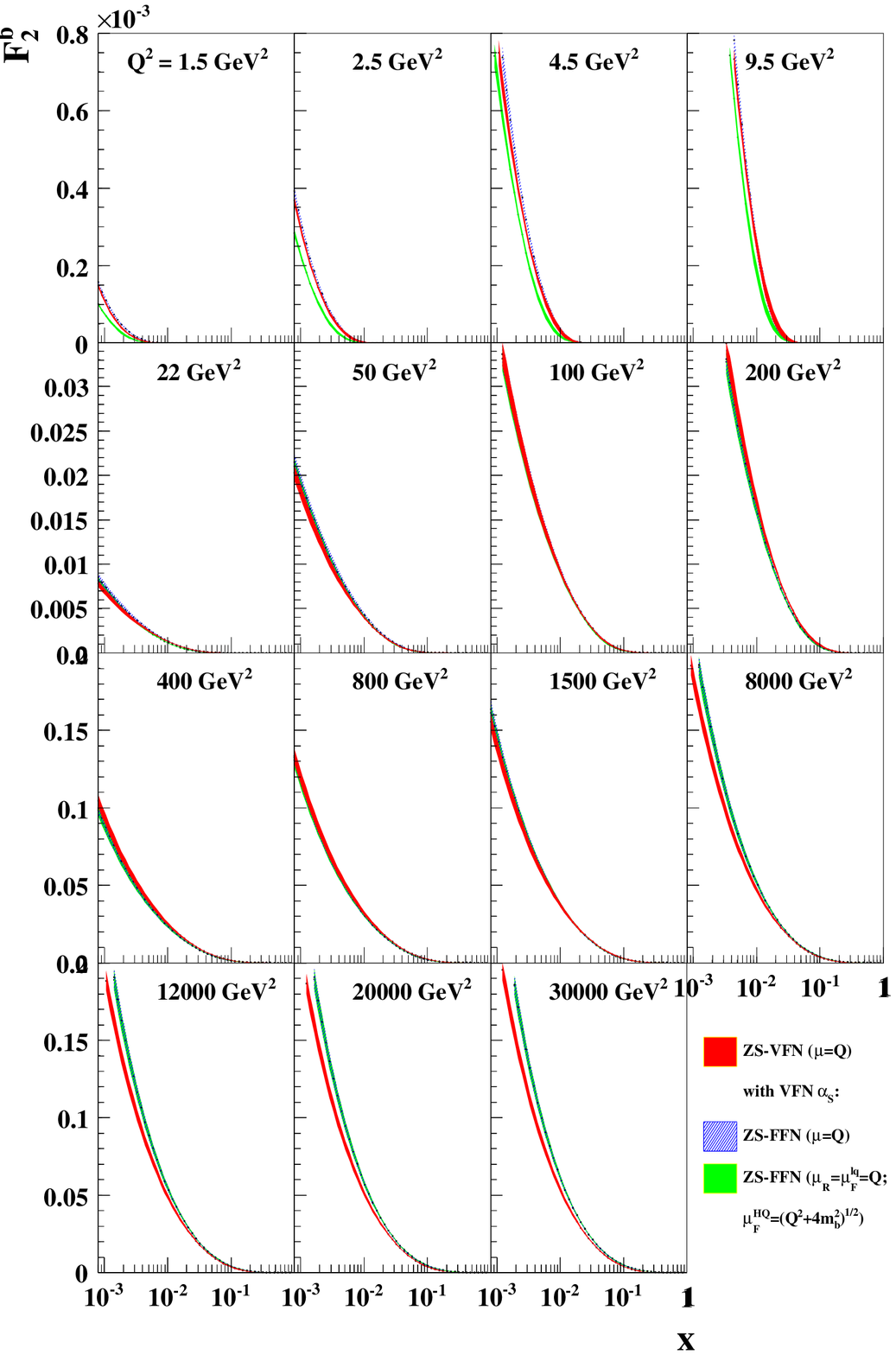,width=0.5\textwidth,height=6cm}
\epsfig{figure=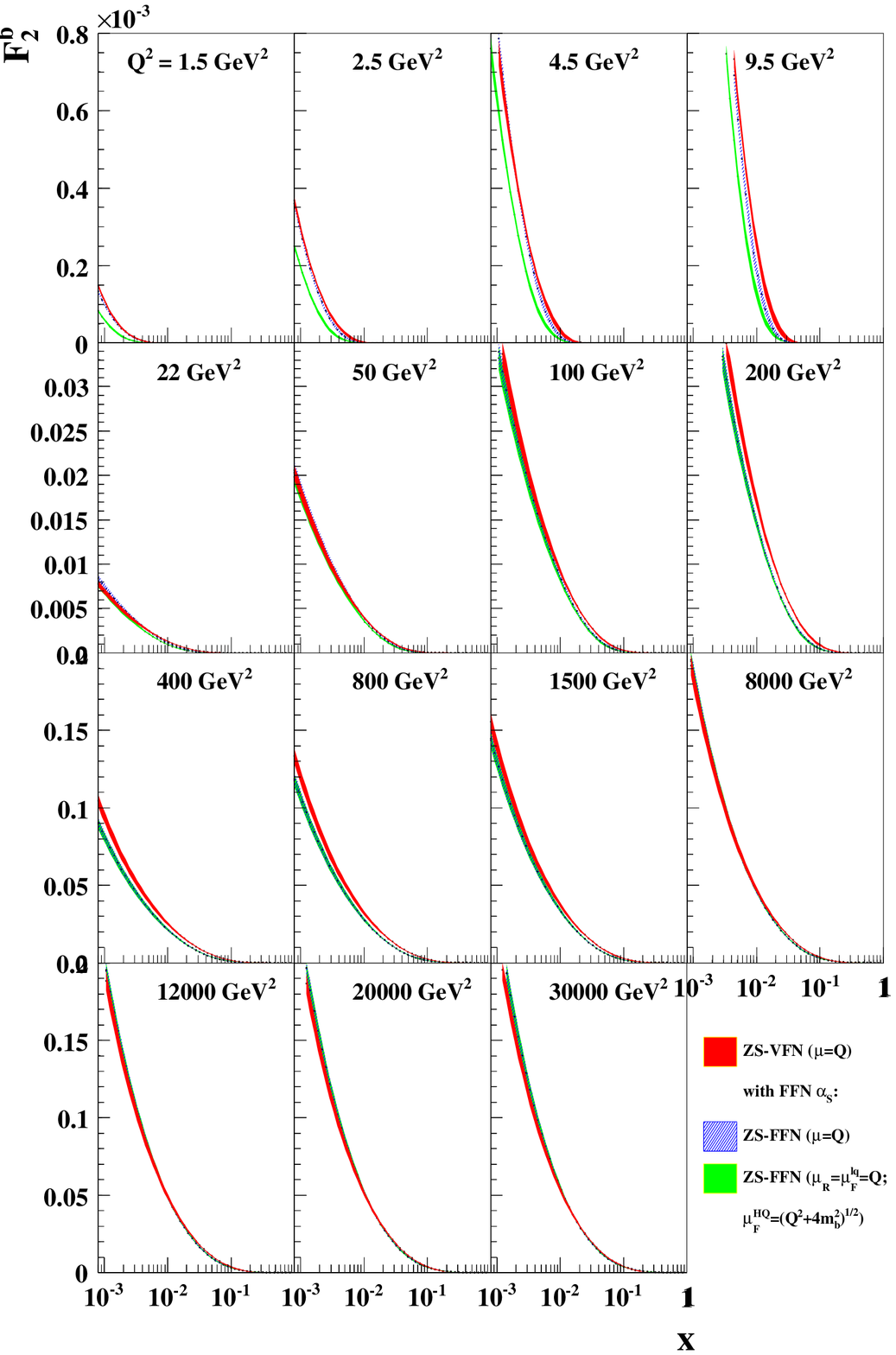,width=0.5\textwidth,height=6cm}}
\caption {Comparison of predictions for $F_2^{b\bar{b}}$, from fits which 
use the TR-VFN scheme and the FFN scheme with two different factorisation 
scales: on the left hand side the FFN schemes still use a VFN treatment of
$\alpha_s$, whereas on the right hand side a 3-flavour $\alpha_s$ is used.}
\label{fig:f2balphas}
\end{figure}

In Fig~\ref{fig:f2calphas} we compare different heavy quark factorisation 
scales and different treatments of the running of 
$\alpha_s$ for predictions of $F_2^{c\bar{c}}$. 
Fig.~\ref{fig:f2balphas} makes the same comparision for 
$F_2^{b\bar{b}}$~\footnote{Note the 
predictions are always made by refitting PDF parameters for each 
scheme choice, not by simply changing the scheme with the same PDF parameters}
.
We see that within the FFN scheme the choice of the heavy quark factorisation 
scale makes only a small difference at low $Q^2$. 
The treatment of $\alpha_S$ gives larger differences. The FFN scheme and 
TR-VFN scheme differ for almost all $Q^2$ if $\alpha_S$ runs as for the VFN 
schemes. However if a 3-flavour $\alpha_S$ is applied in the FFN schemes there 
is much better agreement of all schemes at higher $Q^2$.

\begin{figure}[tbp]
\centerline{
\epsfig{figure=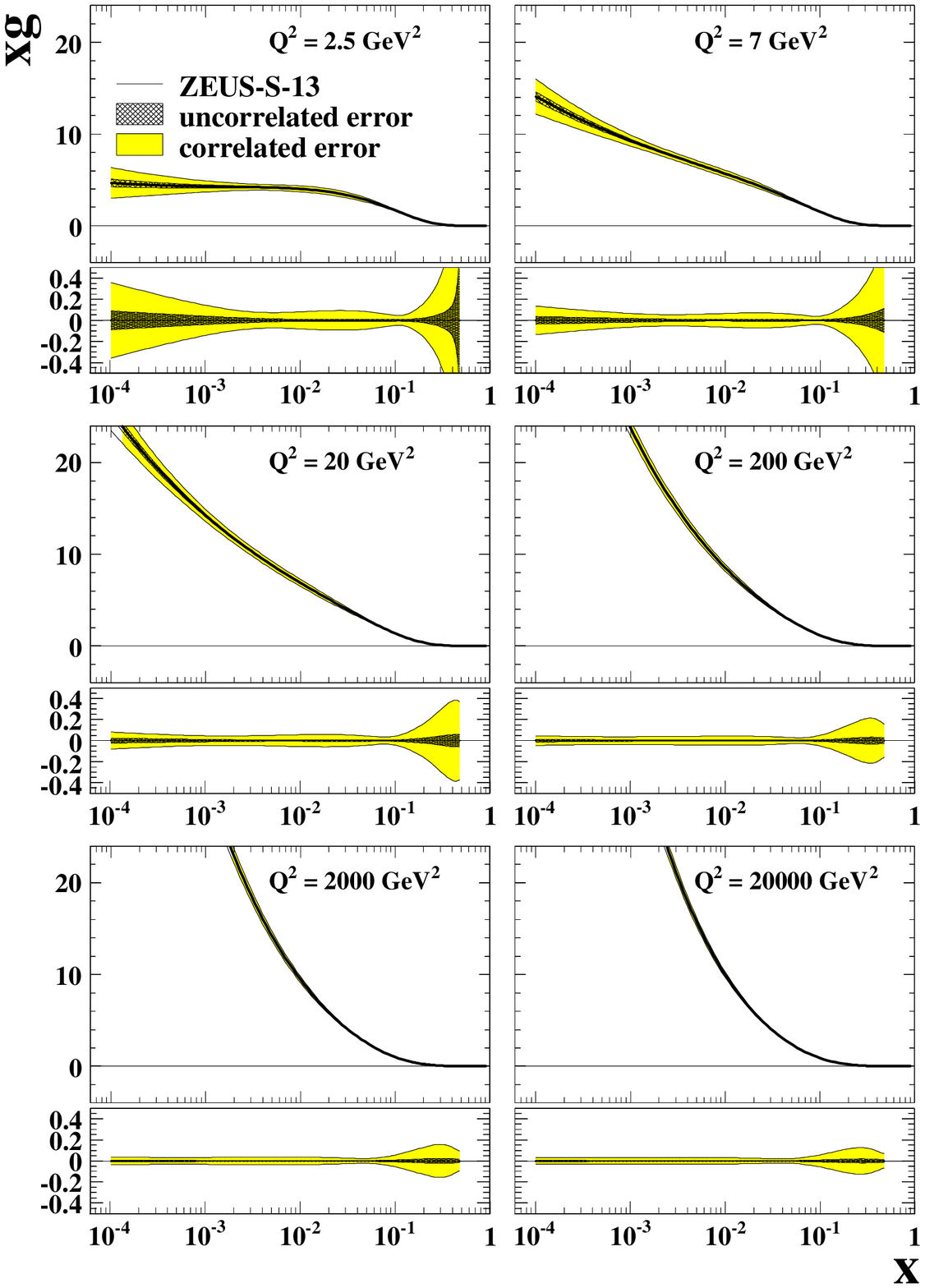 ,width=0.5\textwidth,height=8cm}
\epsfig{figure=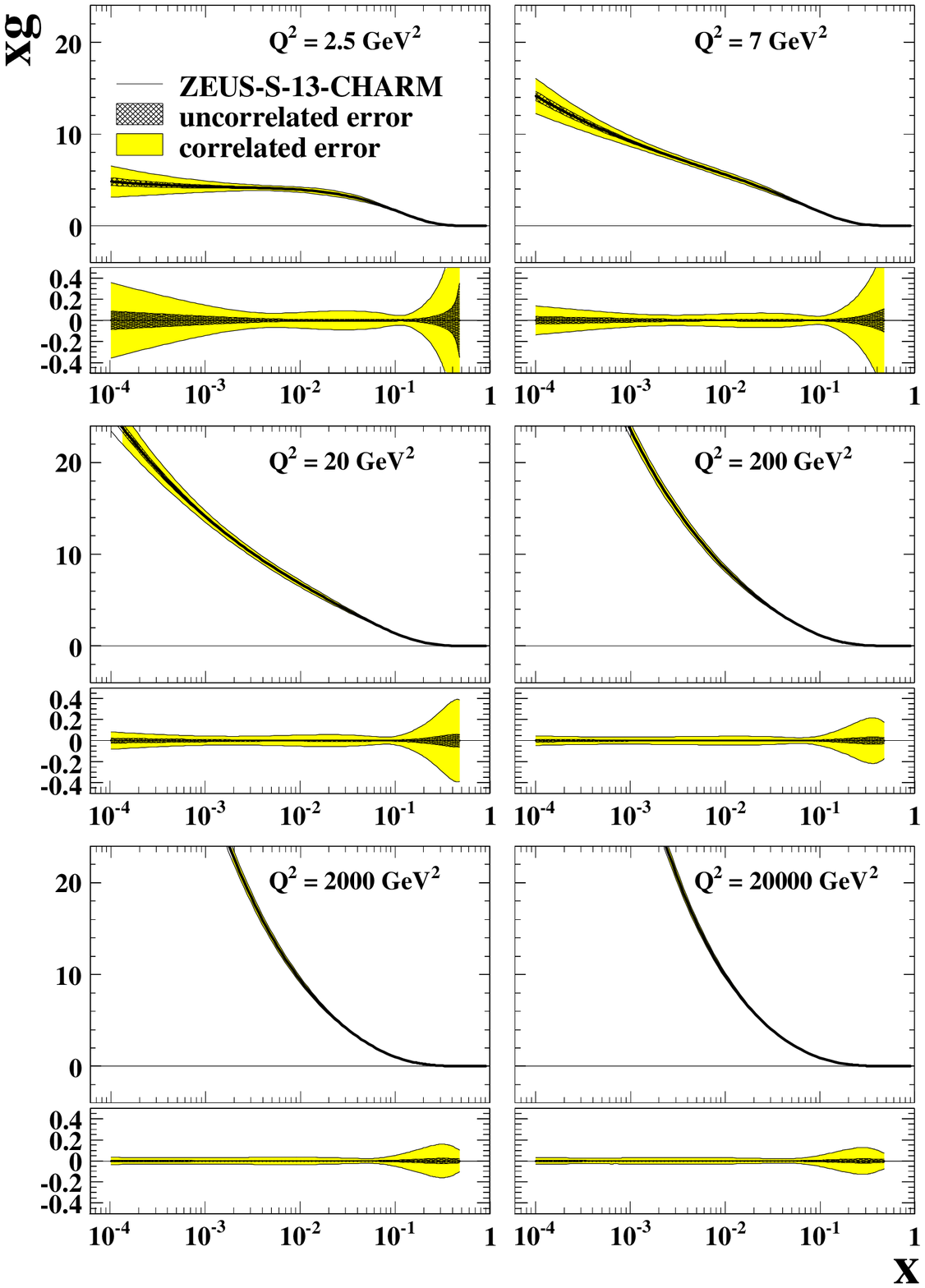 ,width=0.5\textwidth,height=8cm}}
\vspace{-0.5cm}
\caption {The gluon PDF and its fractional uncertainties for various $Q^2$ bins
Left: before $D^*$ cross-section data are input to the ZEUS-S-13 fit. Right: after $D^*$ cross-section data are input to the ZEUS-S-13 fit}
\label{fig:gluonffnzs13}
\end{figure}

We now return to consider inputting the $D^*$ cross-sections to the PDF fit.
The ZEUS-S global fit formalism is used including 
all ZEUS inclusive neutral and charged current cross-section data from HERA-I and the fixed target data.
The parametrisation was also modified to free the mid-$x$ 
gluon parameter $p_5(g)$ and the low-$x$ valence parameter $p_2(u)= p_2(d)$.
This fit is called ZEUS-S-13.
Figure~\ref{fig:gluonffnzs13} 
shows the difference in the gluon PDF uncertainties, before and afer the 
$D^*$ cross-sections were input to the ZEUS-S-13 global fit. 
Disappointingly the uncertainty on the gluon is NOT much improved.

\begin{figure}[tbp]
\vspace{-2.0cm} 
\centerline{
\epsfig{figure=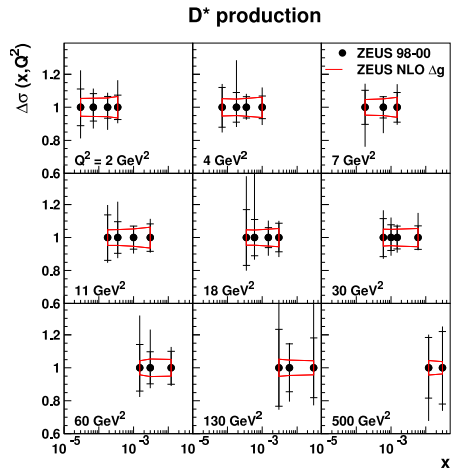,height=8cm}
}
\vspace{-2.0cm}
\caption {Fractional uncertainties on Ddouble differential cross-sections 
for $D^*$ production. the red lines
 show the uncertainties onn these cross-sections deriving from the uncertainty 
on the gluon PDF in the ZEUS-S-13 fit, before including these $D^*$ data in 
the fit.}
\label{fig:d*errors}
\end{figure}

Should we have expected much improvement? 
There are two aspects of the fit which could be improved. The predictiond for 
the $D^*$ cross-sections have more uncertainties than just the PDF 
parametrization. A further  uncertainty is introduced in the choice of the 
$c \to D^*$ fragmentation  The Petersen fragmentation function was used for 
the fit predictions. However, looking back at Fig~\ref{fig:9xsecns} we can 
see that the Lund fragmentation function seems to describe the data better.
To best exploit the charm data in future we need to address such aspects 
of our model uncertainty. Secondly, Fig~\ref{fig:d*errors} compares the 
fractional errors on the $D^*$ cross-sections with the uncertainty on the 
prediction for these quantities derived from the uncertainty on the gluon PDF 
in the ZEUS-S-13 PDF fit, before inputting the $D^*$ cross-sections. The data 
errors are larger than the present level of uncertainty. Thus we 
eagerly await the 5-fold increase in 
statistics expected from HERA-II charm and beauty data.


\section{Acknowledgements}
I would like to thank members of the ZEUS collaboration who have worked on the 
inclusion of heavy flavour data in the PDF fits, particularly: C Gwenlan, 
E Tassi, J Terron, M Wing, M Botje. I thank R Thorne and P thompson for 
useful discussions.

\bibliographystyle{heralhc} 
{\raggedright
\bibliography{heralhc}
}
\end{document}